\documentclass[conference]{IEEEtran}
\IEEEoverridecommandlockouts

\ifCLASSINFOpdf
\else
\fi
\usepackage{srcltx}
\usepackage{graphicx}

\begin{document}
%
% paper title
% can use linebreaks \\ within to get better formatting as desired
\title{Exploring shot noise and Laser Doppler imagery with heterodyne holography}

% author names and affiliations
% use a multiple column layout for up to three different
% affiliations
\author{\IEEEauthorblockN{Michel Gross, Fr\'{e}d\'{e}ric Verpillat, Fadwa Joud}
\IEEEauthorblockA{Laboratoire Kastler Brossel \\UMR 8552 CNRS\\
Ecole Normale Sup\'{e}rieure\\Universit\'{e} Pierre et Marie Curie\\
24, rue Lhomond 75231 Paris Cedex 5 (France)\\
Email: gross@lkb.ens.fr \\verpillat@lkb.ens.fr\\joud@lkb.ens.fr} \and \IEEEauthorblockN{Michael Atlan}
\IEEEauthorblockA{Institut Langevin\\ UMR7587 CNRS; U979 INSERM\\ ESPCI ParisTech\\ Universit\'{e} Pierre et Marie
Curie\\Fondation Pierre-Gilles de Gennes\\10, rue Vauquelin 75005 Paris (France)\\
Email: atlan@optique.espci.fr} }

% conference papers do not typically use \thanks and this command
% is locked out in conference mode. If really needed, such as for
% the acknowledgment of grants, issue a \IEEEoverridecommandlockouts
% after \documentclass

% for over three affiliations, or if they all won't fit within the width
% of the page, use this alternative format:
%
%\author{\IEEEauthorblockN{Michael Shell\IEEEauthorrefmark{1},
%Homer Simpson\IEEEauthorrefmark{2},
%James Kirk\IEEEauthorrefmark{3},
%Montgomery Scott\IEEEauthorrefmark{3} and
%Eldon Tyrell\IEEEauthorrefmark{4}}
%\IEEEauthorblockA{\IEEEauthorrefmark{1}School of Electrical and Computer Engineering\\
%Georgia Institute of Technology,
%Atlanta, Georgia 30332--0250\\ Email: see http://www.michaelshell.org/contact.html}
%\IEEEauthorblockA{\IEEEauthorrefmark{2}Twentieth Century Fox, Springfield, USA\\
%Email: homer@thesimpsons.com}
%\IEEEauthorblockA{\IEEEauthorrefmark{3}Starfleet Academy, San Francisco, California 96678-2391\\
%Telephone: (800) 555--1212, Fax: (888) 555--1212}
%\IEEEauthorblockA{\IEEEauthorrefmark{4}Tyrell Inc., 123 Replicant Street, Los Angeles, California 90210--4321}}

% use for special paper notices
\IEEEspecialpapernotice{(Invited Paper)}

% make the title area
\maketitle

\begin{abstract}
Heterodyne Holography  is a variant of Digital Holography, where the optical frequencies of signal and reference arms can be freely adjusted by acousto-optic modulators. Heterodyne Holography is an extremely versatile and reliable holographic technique, which is able the reach the shot noise limit in sensitivity at very low levels of signal. Frequency tuning enables Heterodyne Holography to become a Laser Doppler imaging technique that is able to analyze various kinds of motion.
\end{abstract}

% IEEEtran.cls defaults to using nonbold math in the Abstract.
% This preserves the distinction between vectors and scalars. However,
% if the conference you are submitting to favors bold math in the abstract,
% then you can use LaTeX's standard command \boldmath at the very start
% of the abstract to achieve this. Many IEEE journals/conferences frown on
% math in the abstract anyway.

% no keywords

% Please choose the correct copyright notice. \IEEEpubidadjcol must be issued somewhere in the second column of the
%title page. This is needed because LATEX resets the text height at the beginning of each column. \IEEEpubidadjcol "pulls
%up" the text in the second column to prevent it from blindly running into the publication ID.
\IEEEpubid{
%For papers in which all authors are employed by the US government, the notice is:
%{U.S. Government work not protected by U.S. copyright}
%For papers in which all authors are employed by a Crown government (UK, Canada, and Australia), the notice is:
%{978-1-4244-8227-6/10/\$26.00~\copyright~2010 Crown}
%For all other papers the notice is:
{978-1-4244-8227-6/10/\$26.00~\copyright~2010 IEEE}
}

%}
% For peer review papers, you can put extra information on the cover
% page as needed:
% \ifCLASSOPTIONpeerreview
% \begin{center} \bfseries EDICS Category: 3-BBND \end{center}
% \fi
%
% For peerreview papers, this IEEEtran command inserts a page break and
% creates the second title. It will be ignored for other modes.
\IEEEpeerreviewmaketitle

\section{Introduction}

Heterodyne Holography (HH) \cite{Leclerc2000,leclerc2001sae} is a variant of the Digital Holography technique (i.e.
holography where the holographic film is replaced by a Digital camera) where the frequencies of the illumination beam
(signal arm) and of the  holographic reference beam (local oscillator arm) are freely adjusted thanks to Acousto-Optic
Modulators (AOM). This technique is called Heterodyne Holography, because it can be viewed  as an optical heterodyne
detection, in which the detector is a multipixel detector  (i.e. a camera) which acquires the heterodyne signal on all
the pixels at the same time keeping trace of the pixel to pixel correlations. These correlations bring the holographic
information that can be used for holographic reconstruction.

\section{Heterodyne holography}

\begin{figure}[h]
\begin{center}
 % Requires \usepackage{graphicx}
  \includegraphics[width=8.5cm]{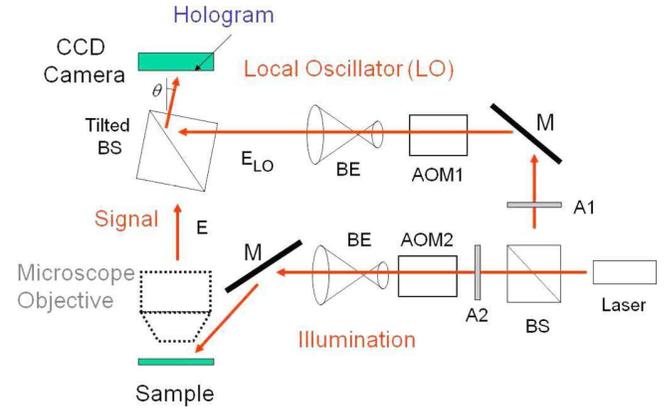}\\
  \caption{Digital holography setup. AOM1 and AOM2: Acousto-optic modulators;
BS: Beam splitter; BE: Beam expander; M: Mirror; A1 and A2: Attenuator; $\theta$ : Tilt angle of the beam splitter with
respect to optical axis.}\label{Fig_Fig1_setup}
\end{center}
\end{figure}

%Heterodyne Holography (HH) \cite{Leclerc2000,leclerc2001sae} is  variant of Digital Holography technique (i.e.
%holography where the holographic film is replaced by a Digital camera) where the frequencies of the illumination beam
%(signal arm) and of the  holographic reference beam (local oscillator are) are freely adjusted thanks to Acousto Optic
%Modulators (AOM). This technique is called Heterodyne Holography, because it can be viewed  as an optical heterodyne
%detection, in which the detector is a multipixel detector  (i.e. a camera) which acquires the heterodyne signal on all
%the pixels at the same time keeping trace of the pixel to pixel correlations. These correlations bring the holographic
%information that can be used for holographic reconstruction.

A typical heterodyne holographic setup is presented on Fig.\ref{Fig_Fig1_setup}. Thanks AOM1,and AOM2, the frequency
$\omega_{LO}$ of the local oscillator field $E_{LO}$ and the frequency $\omega$ of the signal field $E$ are
\begin{eqnarray}\label{Eq_omega_LO}
% \nonumber to remove numbering (before each equation)
\omega_{LO}&=&\omega_L + \omega_{AOM1}\\
\nonumber
 \omega&=&\omega_L + \omega_{AOM2}
\end{eqnarray}
where  $\omega_{L}$ is the main laser frequency, and $ \omega_{AOM1,2}\simeq 80 $MHz are the frequencies of the signals that drive AOM1 and AOM2.

\newpage

This setup  is extremely versatile and allows to perform either off-axis holography \cite{schnars2002digital} or phase
shifting holography \cite{Yamaguchi1997}. For example, to perform a 4-phase detection, the frequency shift between the LO and signal beam must be chosen such as:
\begin{eqnarray}\label{Eq_omega_4phases}
% \nonumber to remove numbering (before each equation)
\omega_{LO}- \omega = \omega_{AOM1}- \omega_{AOM1}=\omega_{CCD}/4
\end{eqnarray}
where $\omega_{CCD}$ is the frame rate of the camera. In that case, the holographic signal $E E_{LO}^*$ is obtained by
4 phases demodulation of the camera signal. If $I_n$ where $n=0...3$ is the camera signal, and $n$ camera frame index,
the holographic signal $E E_{LO}^*$ is then given by:
\begin{eqnarray}\label{Eq_omega_4phases_E_ELO}
E E_{LO}^* = (I_0-I_2)+ j (I_1-I_3)
\end{eqnarray}
where $j^2=-1$.
\newpage

Here, by combining the frames with the 4-phase linear combination $(I_0-I_2)+ j (I_1-I_3)$ of
Eq.\ref{Eq_omega_4phases}, we tune the heterodyne receiver at frequency
\begin{eqnarray}\label{Eq_omega_H}
\omega_{H}=\omega_{LO}-\omega_{CCD}/4
\end{eqnarray}
Because of Eq.\ref{Eq_omega_4phases}, $ \omega_{H}$ is
equal to $\omega$, and the heterodyne receiver is tuned at
the signal frequency $\omega$. One of the advantage of
method, used here  to shift the phase (AOMs), is the
accuracy of the $\pi/2$ phase shift that is applied to the
reference beam, from one frame to the next
\cite{atlan2007accurate}. As a consequence, the parasitic
twin image \cite{cuche2000spatial} is minimized.

We must note that it is is possible to make another choice for the frequency offset $\omega_{LO} - \omega$. In that
case, the holographic detector will be able to explore different frequency components, opening the way to Laser Doppler Holographic Imaging \cite{atlan2006laser_Doppler}, and to Sideband Digital Holographic Imaging \cite{joud2009imaging}.

\section{Shot Noise Holography}

Because the holographic signal results from  the
interference of the object complex field $E$ with a
reference complex field $E_{LO}$, whose amplitude can be
much larger (i.e. $E_{LO}\gg E$), it benefits of an optical
heterodyne gain
\begin{eqnarray}\label{Eq_G_H}
G_H = \frac{|E E^*_{LO}|}{|E|^2}
\end{eqnarray}
and is thus potentially well-suited  for the detection of
signal fields $E$ of weak amplitude
\cite{charrirere2007influence,charriere2006shot}.

To better analyze this effect,  one must measure the camera
signal in photoelectron (e) Units. In typical experimental
conditions, the LO beam intensity is adjusted to be at half
saturation of the camera. With a 12 bits camera, whose gain
is $G \simeq 5 $ e/DC  (photoelectron per Digital Count),
the LO beam signal is then
\begin{eqnarray}\label{Eq_I_n_in_e_units}
I_n\simeq |E_{LO}|^2 \sim 2000~\textrm{DC} \simeq 10^4~\textrm{e}
\end{eqnarray}
per pixel, and the LO beam  shot noise standard deviation
is  $\sqrt{I_n} \sim 100$.

For a weak signal of about  1 e per pixel,  the heterodyne
gain is $G \sim 100$. The 1 e heterodyne signal:  $|E
E^*_{LO}| \sim 100$ e, is thus equal to the LO beam shot
noise, yielding SNR =1 (Signal to Noise Ratio). We must
note here that the camera reading noise ($\simeq 20$ e),
and the camera Analog Digital Converter quantization noise
($\simeq 5$ e) can be neglected with respect to the LO beam
shot noise ($\sim 100$ e) \cite{gross2008noise}.

\begin{figure}[h]
\begin{center}
 % Requires \usepackage{graphicx}
  \includegraphics[width=8.5cm]{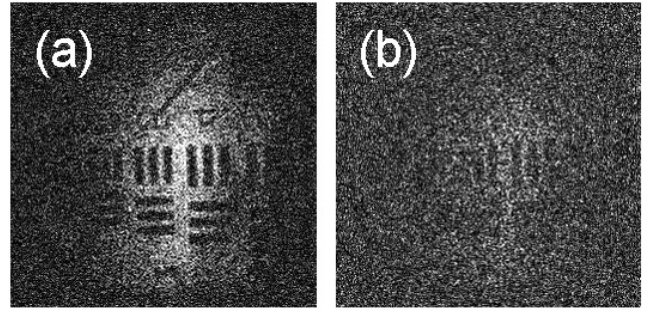}\\
  \caption{
Reconstructed image of a USAF target from a  sequence of
$M=12$ images.  The size of calculation grid  (which is
equal to the size of the recorded hologram) is $A_{calc} =
1024 \times 1024 $ pixels. The size of the spatial filter
\cite{cuche2000spatial} selected zone is $ A_{calc}=400
\times 400 $ pixels. Average signal for $|E|^2$ is
$\eta/M=0.0127$ (a), and  $0.1 \eta/M=0.00127$ e per pixel
and per frame in (b), corresponding to $SNR \simeq 1$, and
0.1 respectively. }\label{Fig_Fig2_USAF}
\end{center}
\end{figure}

One can generalize this result for a sequence  of $M$
images. One get that the shot noise limit, which
corresponds to SNR =1, is reached for 1 e per pixel for the
whole sequence of $M$ images, i.e. for $1/M$ e per pixel
and per frame. One must notice also that the total amount
of shot noise is proportional to the number of pixels (or
spatial modes) that are used in the reconstruction
procedure. If spatial filtering in the Fourier space is
used \cite{cuche2000spatial}, the shot noise must be
multiplied by the ratio
\begin{eqnarray}\label{Eq_eta}
\eta = A_{filter}/A_{calc}<1
\end{eqnarray}
where $A_{filter}$ is the number of pixels  of the spatial
filter in Fourier space, and $A_{calc}$ the number of
pixels of the calculation grid. The shot noise limit
corresponds thus to $\eta/M$ e per pixel and per frame
\cite{gross2007digital}.

Figure \ref{Fig_Fig2_USAF} shows, as an example,  images of
an USAF target that has been reconstructed from holographic
data recorded at very low signal levels ($0.0127$ and
$0.00127$ e per pixel and per frame in (a) and (b)
respectively). These images illustrate the ability of
heterodyne holography to reach the shot noise limit in low
light with a standard array detector.

\section{Laser Doppler Holography}

\subsection{Flows}

\begin{figure}[h]
\begin{center}
 % Requires \usepackage{graphicx}
  \includegraphics[width=4.2cm]{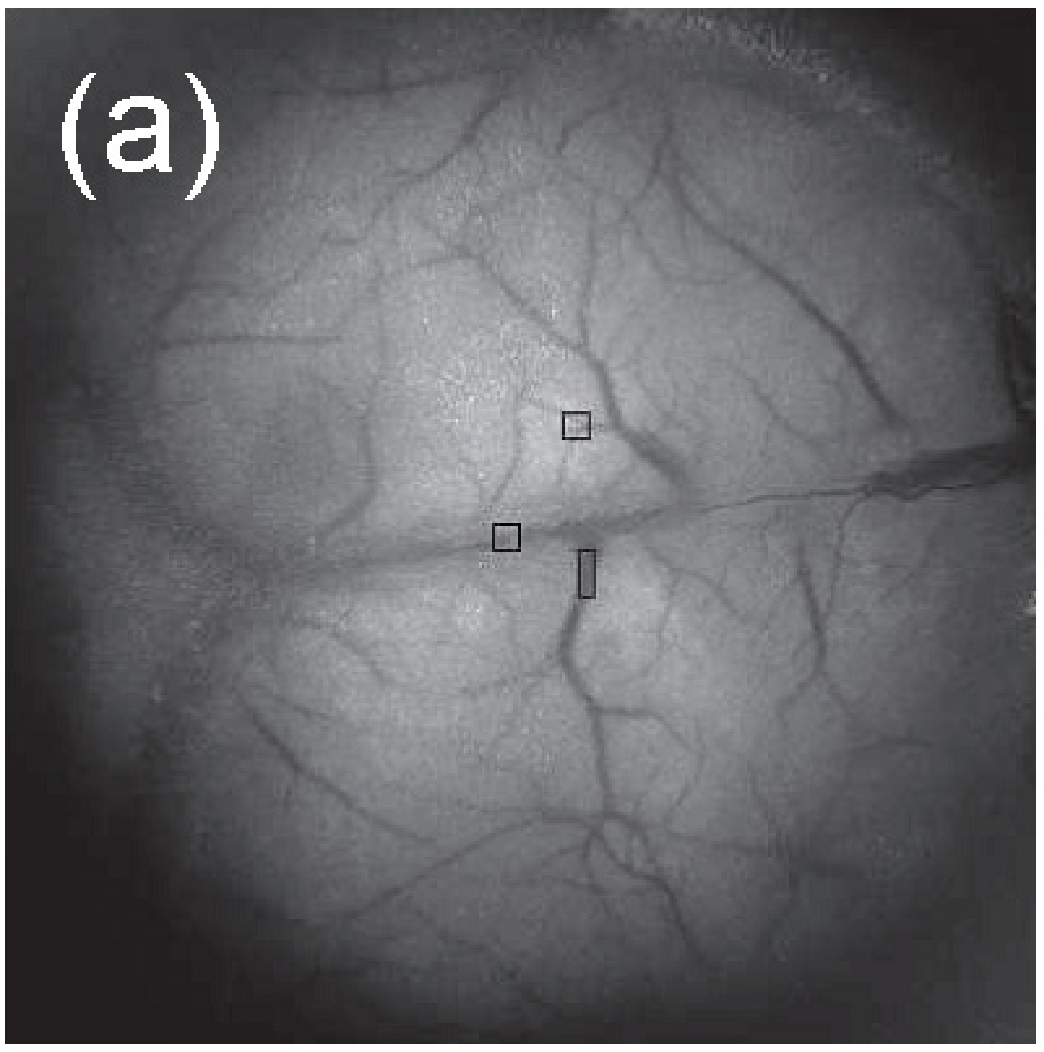}
  \includegraphics[width=4.2cm]{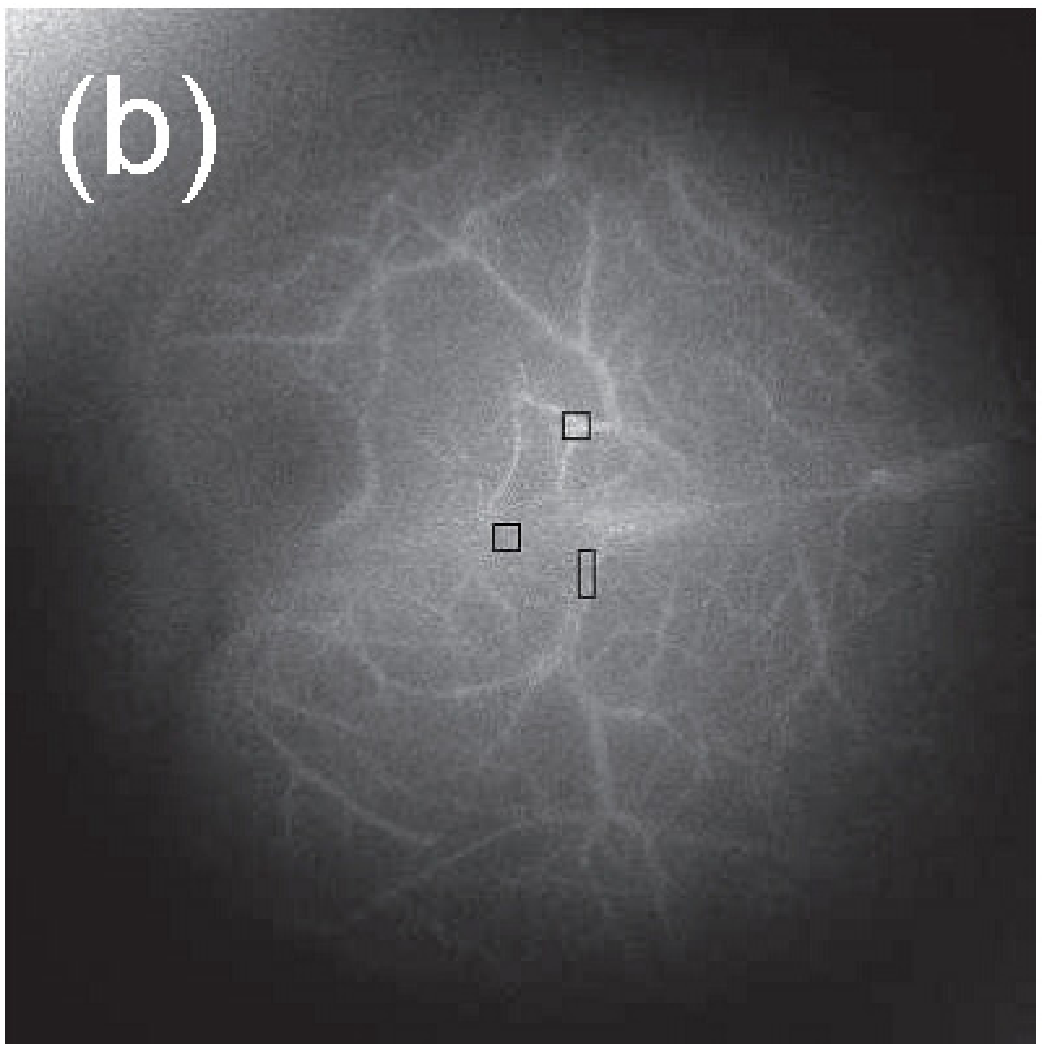}\\
  \caption{
Image of a mouse crania reconstructed by Laser Doppler Holography: average of the intensity images reconstructed from
data recorded with $\omega_{LO}-\omega$ varying from 36 to 48 Hz (a), and from 0.6 to 2.0 kHz (b).
}\label{Fig_Fig_mouse}
\end{center}
\end{figure}

The ability to control  the frequency of the heterodyne
detection $\omega_{H}$, following Eq.\ref{Eq_omega_H},
allow us to develop a Laser Doppler Holography. Thanks to
the AOMs, we can adjust the frequency of the local
oscillator $\omega_{LO}$ so as we can select a velocity
component, whose optical frequency is $\omega_{H}$, and we
can image it  by holography \cite{atlan2006laser_Doppler}.
It is also possible to reconstruct the images of the
various velocity components by recording a sequence of
images ($I_1, I_2 .....I_n$), and by making a time to
frequency Fourier transformation
\cite{atlan2007wide,atlan2008high}.

As the optical wavelength is much  smaller than the
ultrasound wavelength, this  Laser Doppler Holography
presents the advantage to allow the analyze flows in the
$\mu m/s$ range \cite{atlan_microflow_2006}. This Laser
Doppler Holography has been applied to the analysis of
blood flow in capillaries. We have for example imaged
cerebral blood flow in the micro vessels, in mice, in vivo
\cite{atlan2006frequency,atlan2007cortical,atlan2008high}.
The imaging is made in vivo, in minimally invasive
conditions (the cranial skull remains present). Figure
\ref{Fig_Fig_mouse} gives an example of mouse cerebral
blood flow images that can be obtained by this way. Similar
results have been obtained on the retina of a rat
\cite{simonutti2010holographic}.

\subsection{Vibrations}

\begin{figure}[h]
\begin{center}
 % Requires \usepackage{graphicx}
  \includegraphics[width=8.5cm]{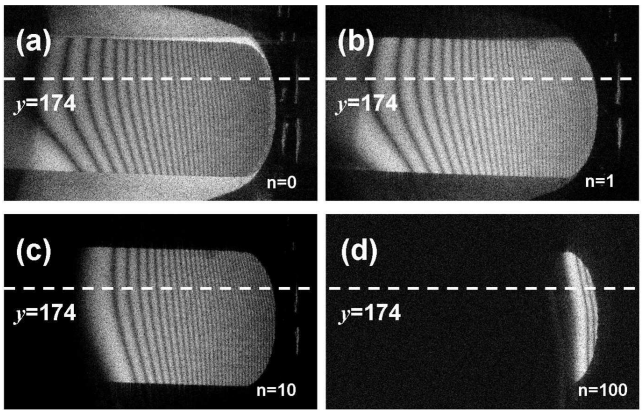}\\
  \caption{
Reconstructed holographic images of a clarinet reed vibrating at frequency $ \omega_v= 2143$ Hz perpendicularly to the
plane of the figure. Figure (a) shows carrier image obtained with  $\omega_H=\omega$ ; Fig. (b)-(d) show the frequency
sideband images obtained with $\omega_H=\omega + n \omega_v $, where $n$ the sideband harmonic order: $n = 0$ (a), 1
(b), 10 (c) and  100 (d). }\label{Fig_Fig_mouse}
\end{center}
\end{figure}

If the object vibrates periodically , the light that is reflected or diffracted
by the object exhibit sidebands. It is then possible to tune the holographic
heterodyne receiver at a sideband frequency \cite{Aleksoff_71,joud2009imaging},
and not simply at the carrier frequency like with standard digital holography
\cite{picart2003time}. If the amplitude of vibration is large, it is even
possible to explore the whole comb of sideband lines, and to get, by the way,
the map of the vibration amplitudes of the object. Figure \ref{Fig_Fig_mouse}
shows sideband images of a vibrating clarinet reed. Images with harmonic order
up to $n\simeq 1200$ have been obtained by Heterodyne Holography
\cite{joud2009fringe}.

\section{Conclusion}

Heterodyne Holography is an extremely versatile and robust holographic
technique,  which is able to reach the shot noise limit of sensitivity in low
light, and that is perfectly suited to the analysis of many modulation
phenomena, flows or vibrations  \cite{atlan2007cortical,colomb2006total,kuhn2009submicrometer}.

%
%\section*{Acknowledgment}
%

\bibliographystyle{IEEEtran}
%\bibliography{wio__10}

% that's all folks
\end{document}